\documentclass[journal]{IEEEtran}
\usepackage{graphicx}
\usepackage{multirow}
\usepackage{graphicx}
\usepackage{graphicx}
\usepackage{amsmath}
\usepackage{subfigure}
\usepackage{rotating}
\usepackage{multirow}
\usepackage{array}
\usepackage{algorithm}
\usepackage{algorithmic}
\usepackage{cases}
\usepackage{mathtools}

\begin{document}

\title{Performance Analysis of Hierarchical Routing Protocols in Wireless Sensor Networks}

\author{\IEEEauthorblockN{K. Latif, M. Jaffar, N. Javaid, M. N. Saqib, U. Qasim$^{\ddag}$, Z. A. Khan$^{\$}$\\\vspace{0.4cm}}

        $^{\ddag}$University of Alberta, Alberta, Canada.\\
        Department of Electrical Engineering, COMSATS\\ Institute of
        Information Technology, Islamabad, Pakistan \\
        $^{\$}$Faculty of Engineering, Dalhousie University, Halifax, Canada.

     }

% make the title area
\maketitle

\begin{abstract}
\boldmath
This work focusses on analyzing the optimization strategies of routing protocols with respect to energy utilization of sensor nodes in Wireless Sensor Network (WSNs). Different routing mechanisms have been proposed to address energy optimization problem in sensor nodes. Clustering mechanism is one of the popular WSNs routing mechanisms. In this paper, we first address energy limitation constraints with respect to maximizing network life time using linear programming formulation technique. To check the efficiency of different clustering scheme against modeled constraints, we select four cluster based routing protocols; Low Energy Adaptive Clustering Hierarchy (LEACH), Threshold Sensitive Energy Efficient sensor Network (TEEN), Stable Election Protocol (SEP), and Distributed Energy Efficient Clustering (DEEC). To validate our mathematical framework, we perform analytical simulations in MATLAB by choosing number of alive nodes, number of dead nodes, number of packets and number of CHs, as performance metrics.
\end{abstract}


\IEEEpeerreviewmaketitle
\section{Introduction}
\IEEEPARstart{W}{ireless} Sensor Networks (WSNs) are composed of tiny and miniaturized electronic devices which are known as sensors. Sensors can sense, compute, store, transmit and receive data of interests from the environment in which they are deployed. Due to small size of sensors, a big size battery source can not be embedded into them therefore sensors need efficient mechanism for energy utilization. To improve  the life time of the sensors in WSNs, communication protocols plays an important role. The design objective of these protocol is avoiding unnecessary data transmission and reception. For this purpose, switching the nodes into idle or sleep mode when there is no data to send or receive. For efficient utilization of energy resources, many routing and Medium Access Control (MAC) layer protocols are defined.

A Sensor Node (SN) is composed of processor, sensor, transceiver, and power units. In addition to performing these functionalities, a sensor node also has the capability of routing. Due to the remote nature of WSNs deployment, sensor nodes face energy optimization and quick route discovery problems, different routing techniques have been proposed to address these issues. Clustering is one of them and is used in WSNs, which handles these issues efficiently.
Base Station (BS) and Cluster Head (CH) are the main components of a clustered network. In a cluster, SNs are located at minimum communication distance. Each cluster is headed by a CH. All the nodes in a cluster are in accessible range of the CH. Member nodes in a cluster send their data to their respective CH, and CH aggregates data and sends aggregated data to the BS. In this paper we first address energy limitation constraint with respect to maximizing network life with respect to bandwidth consumption. We select different clustering mechanisms, i.e., Low Energy Adaptive Clustering Hierarchy (LEACH) [1], Threshold Sensitive Energy Efficient sensor Network (TEEN) [2], Stable Election Protocol(SEP) [3], and Distributed Energy Efficient Clustering (DEEC) [4], how these clustering mechanisms provide feasible solution against modeled constraints.

To validate our mathematical framework, we perform analytical simulations in MATLAB. Different performance parameters; number of alive nodes, number of dead nodes, number of packets and number of CHs are selected for this purpose. For evaluating routing strategies, based on clustering techniques in literature different clustering techniques have been introduced. Almost all of the clustering techniques consist of two phases, i.e., setup phase and steady state phase. In setup phase, election of CH and formation of cluster is performed, while in steady state phase data is transmitted from node to CH, CH then aggregates this data and transmit it to BS.

\begin{table*}[t]
\caption{Detailed Comparison of Clustering Protocols in the Heterogeneous Environment}
\centering
\begin{tabular}{|c|p{1.75cm}|p{1.5cm}|p{1.5cm}|p{1.5cm}|p{1.5cm}|p{1.5cm}|p{1.5cm}|p{1.5cm}|p{1.5cm}|}
\hline\bf{Protocol}&\bf{Initial Energy Level}&\bf{Scalability}&\bf{Energy Efficiency}&\bf{Network Lifetime}&\bf{Data Aggregation}& \multicolumn{3}{|c|}{\bf{CH Selection Criteria}}\\ \cline{7-9} &&&&&& Initial Enrgy& Residual Enrgy& Avg. Net. Enrgy\\ \hline
LEACH&Single level&Limited&Low&Poor&Yes&Yes&No&No \\ \hline
TEEN&Single level&Limited&Very high&Best&Yes&Yes&No&No \\ \hline
SEP& Two levels heterogeneous&scalable&Low&Good&Yes&No&Yes&No \\ \hline
DEEC&Multi levels heterogeneous&Scalable&High&Better&Yes&No&Yes&Yes \\ \hline
\end{tabular}
\end{table*}

\section{Problem Formulation for Network Life Time in WSNs}
Let a set of sensors; $s_1,...,s_N$, with adjustable sensing ranges are deployed in a network. These sensors are alive for a specific number of rounds, $r_1, r_2, ..., r_K$, where $K$ denotes the index for round number. To maximize network life time, $K$ must be maximized in such a way that each sensor appearing in the sets $r_1, r_2, ..., r_K$ consumes at most $E$ energy ($E$ is the initial energy of the sensor nodes).

Maximizing $K$ is equivalent to maximizing life time of a network. The sensing range of sensors in terms of distance, $d$, determines energy consumtion by the sensor during activation period of sensors. If a sensor participates in more than one set, then the sum of energy spent during network life time has to be at most $E$. There are two ranges, $d_2$ and $d_4$. The energy consumption for transmitting at $d_4$ is more as compared to $d_2$. Let us consider for this example $E = 2$, $e_1 = 0.5$, and $e_2 = 1$, $M$ CHs $c_1, c_2,..., c_M$, and $Z$ sensing ranges $p_1, p_2,..., p_Z$ and the corresponding energy consumption $e_1, e_2,..., e_Z$. Initial relationship between sensor and CH: $\alpha_{izj} = 1$, if sensor $s_i$ with radius $r_z$ covers CH $c_{ij}$. Some indexes are also used which are; $i$, $i_{th}$ sensor, $j$, $j_{th}$ CH, $z$, $z_{th}$ sensing range, and $k$, $k_{th}$ round. Some variables are also defined: $r_k$, boolean variable, for $k = 1..K$; $r_k = 1$ there are still alive nodes are present, otherwise $r_k = 0$, if. $x_{ikz}$ is a boolean variable, for $i = 1..N$, $k = 1..K$, $z = 1,...,Z$; $x_{ikz} = 1$ if sensor $i$ with range $p_z$ is in cover $k$, otherwise $x_{ikz} = 0$.

\begin{eqnarray}
Max\,\,\, r_1 + ... + r_K
\end{eqnarray}
\normalsize

\textbf {Subject To}

\begin{equation*}
\begin{aligned}
\begin{split}
&\sum_{k=1}^{K} ( \sum_{z=1}^{Z} x_{ikz}e_z \le E)\,\,\, \forall{i=1,..., N}\,\,\,\text{(1.a)}\\
&\sum_{z=1}^{Z} x_{ikz}\le c_k \,\,\, \forall{i=1,...,N} \,\,and\,\,\forall{k=1,...,K}\,\,\,\text{(1.b)}\\
&\sum_{i=1}^{N} ( \sum_{z=1}^{Z} x_{ikz}\times \alpha_{izj} \ge r_k)\,\,\, \forall{k=1,...,K}\,\,and\,\,\forall{j=1,...,M}\,\,\,\text{(1.c)}\\
&x_{ikz}\in {0, 1}\,\,and\,\,r_k\in {0,1}\,\,\,\text{(1.d)}
\end{split}
\end{aligned}
\end{equation*}

$K$ represents an upper bound for the number of rounds. The first constraint in eq. 1(a), $\sum_{k=1}^{K} ( \sum_{z=1}^{Z} x_{ikz}e_z \le E$ for any $i = 1..N$, assures that the energy consumed by each sensor $i$ is less than or equal initial  energy of each sensor. In eq. 1(b), the constraint $\sum_{z=1}^{Z} x_{ikz}\le r_k $, for any $i = 1..N$ and $k = 1..K$, guarantees that, if sensor $i$ is part of any round $k$ then exactly one of its $Z$ sensing ranges are set. Whereas, in eq. 1(3), $\sum_{i=1}^{N} ( \sum_{z=1}^{Z} x_{ikz}\times \alpha_{izj} \ge r_k$ for any $k = 1..K$ and $j = 1..M$, guarantees that each CH $C_j$ is covered by each round $r_k$. After problem formulation through linear programming technique, now we discuss the strategies of selecting routing protocols for WSNs to maximize the value of $K$.

\section{Clustering Process in Chosen Protocols}
Basically, in an operation of any clustering protocol, three states occurs in following sequence.

\textit{1) Advertisement state:} In this state, for every protocol, CHs are elected on the basis of different parameters like initial energy of each node, remaining energy of every node, and the total network average energy. When CHs are elected, after that, they advertise their status to their respective member nodes (non-CH which are associated with their respective CH) by using CSMA-CA MAC protocol. Selection of non-CH is based on Receive Signal Strength Indicator (RSSI).

\textit{2) Setup state:} A neighboring CH node, after receiving status of cluster members, allocates TDMA based time slots to the Non-CHs nodes, so that they send data to the CH.

\textit{3)Steady state:} It is the data transmission state, in which non-CHs sensors sense environment and send sensed data to CH, during allocated time slots by the CH. A CH then aggregates the sensed data and after compressing the sensed data, it transmits sensed data to the BS.

The clustering process completes in both the advertisement and set up states. Therefore, first of all, we describe cluster head selection criterion in the hierarchical clustering protocols. After it, in the next subsection, we describe cluster formation process in LEACH, SEP, DEEC and TEEN protocols in detail.

\subsection{CH Selection in Hierarchical Clustering Protocols}
In wireless sensor networks, we choose cluster heads for data aggregation and transmission in such a way that more energy is conserved, as given in eq. 1(a), that each node is restricted to utilization of limited energy which is equal to $E$, as mention in eq. 1(a). With the help of CH selection criterion in different protocols (homogenous or heterogenous) may enhance the stability region and life time, $K$, of the whole network. Therefore, we have studied different CH selection criterion for selected protocols.

\subsubsection{LEACH and TEEN Protocols}
 LEACH and TEEN follow self organizing and adaptive CH selection criteria. In setup phase, CH is elected on the bases of following threshold equation [1].

 \begin{equation}
 T(n) =
  \begin{dcases}
   \frac{P}{1-p(r*mod(1/p))} & \text{if } n  \in G\\
   0 & \text{otherwise }
  \end{dcases}
\end{equation}\newline where, $P$ is the desired number of CHs, $r$ is the current round and $G$ is the set of nodes that have not been CH in the current epoch. Epoch is the number of rounds for a CH, after which again it become eligible to become a CH. Each node generates a random number between $0$ and $1$, if the number is less than the node's threshold, then this sensor node becomes a CH. After the election of CHs, each CH advertises its status using CSMA MAC protocol. Node selects its CH, on the bases of RSSI and link quality of all CHs, existing in range of that node. All nodes send their membership willingness message to the suitable CH, using CSMA MAC. Then CHs schedule all nodes using TDMA for data transmission.
In steady-state phase, each node transmits its data to their respective CH in specific allocated time slots. CH then aggregates data and sends the compressed data to BS. Fig. 1 shows clustering mechanism in LEACH.

\begin{figure}[h]
\centering
\includegraphics [height=6.5cm,width=5cm]{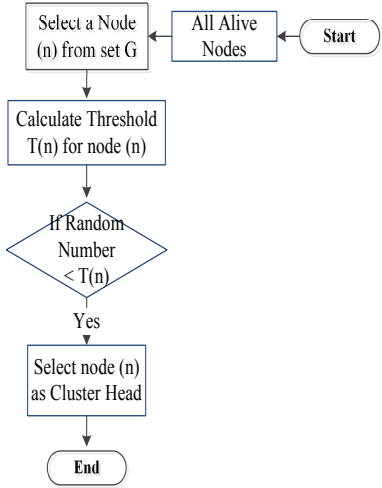}
\caption{Flow chart of CH Selection in LEACH protocol.}
\end{figure}

\subsubsection{SEP Protocol}
SEP is a protocol for heterogeneous network; heterogeneity in terms of initial energy deployment in SNs. SEP assumes that in real environment nodes have different energy, therefore SEP two types of nodes, i.e., advance nodes and normal nodes. Advance nodes have an $\alpha$ amount of more energy than normal nodes. SEP assign a weighted probability to each node based on its initial energy. Moreover, it improves the cluster formation of LEACH by decreasing the CH epoch interval of advance nodes, i.e., advance nodes get more chances to become a CH. LEACH threshold formula in eq. 3 works well for homogeneous energy nodes, however, the problem of maintaining well distributed energy consumption constraints; eq. 1(a-c), in the stable period arises in heterogeneous energy nodes environment. SEP resolve this issue by introducing guaranteed well distributed energy consumption constraint in the stable period [4], for maximizing $K$. For this purpose, a weight is assigned for individual probabilities for election of CHs for advance and normal nodes. Therefore SEP gives two different threshold formulae given in eq. 4 and 6.

 \begin{equation}
 T(S_{nrm}) =
  \begin{dcases}
   \frac{p_{nrm}}{1-p_{nrm}(r*mod(1/p_{nrm}))} & \text{if $S_{nrm}$ $\in$ $G^{'}$}\\
   0 & \text{otherwise }
  \end{dcases}
\end{equation}\newline where, $G^{'}$ is the set of normal nodes which can become CH and

 \begin{equation}
P_{nrm}=\frac{P_{opt}}{1 + a.m}
\end{equation}

 \begin{equation}
 T(S_{adv}) =
  \begin{dcases}
   \frac{p_{adv}}{1-p_{adv}(r*mod(1/p_{adv}))} & \text{if $S_{adv}$ $\in$ $G^{''}$}\\
   0 & \text{otherwise }
  \end{dcases}
\end{equation}\newline where, $G^{''}$ is set of advance nodes, which can become CH and

  \begin{equation}
P_{adv}=\frac{P_{opt}}{1 + a.m} (1 + \alpha)
\end{equation}

\begin{figure}[h]
\centering
\includegraphics [height=8cm,width=7cm]{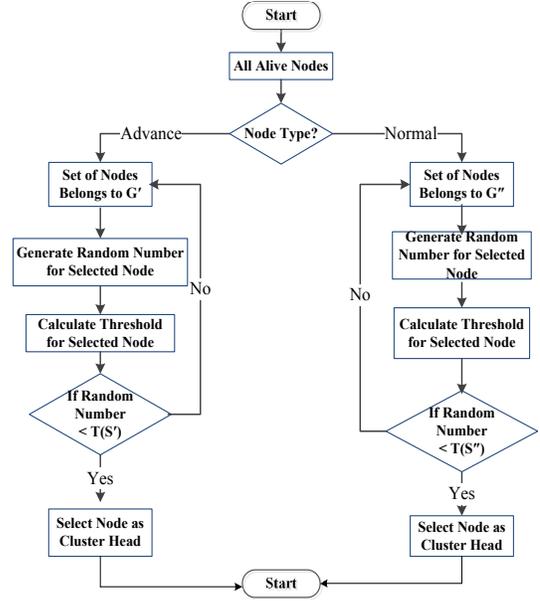}
\caption{Flow chart of CH Selection in SEP protocol.}
\end{figure}
CH selection process of SEP is depicted in Fig. 2.

\subsubsection{DEEC Protocol}
DEEC is another enhancement of LEACH for multi-level heterogeneous environment with respect to level of energies in WSNs. In SEP, energy distribution for two levels, i.e., advance nodes and normal nodes, whereas, DEEC introduces multi-level heterogeneity for maximizing $K$. The nodes having greater residual energy have more right to become a CH. Therefore, CH formation in DEEC is based on residual energy of entire network and residual energy of the node that wants to become a CH. SEP calculates optimum number of CHs from equation 5 and 7 for advance and normal nodes, respectively. While, in DEEC, for multi-level heterogeneous node energy environment, nodes with higher residual energy attains more chances to become a CH. Therefore, DEEC calculate optimum number of CHs for each round from the following two equations [4].

 \begin{equation}
 P(i) =
  \begin{dcases}
   \frac{p_{opt}E_i(r)}{(1+am)E^{'}(r)} & \text{if $S_{i}$ is normal node} \\
   \frac{p_{opt}(1+\alpha)E_i(r)}{(1+am)E^{'}(r)} &\text{if $S_{i}$ is advance node}
  \end{dcases}
\end{equation}\newline where, $E^{'}$(r) is the average energy of the network at round $r$ and is given by [4]:

 \begin{equation}
E^{'}(r)=\frac{1}{N}\sum_{i=1}^{N}E_{i}(r)
\end{equation}\newline $E_{i}(r)$ is the residual energy of the node at round $r$. Based on $P_i$, DEEC calculates threshold [4] as:

  \begin{equation}
P(S_{adv})=\frac{P_{opt}N(1 +  \alpha_{i})}{(N + \sum_{i=1}^{N}\alpha_{i})}
\end{equation}

DEEC evaluates that if the residual energy of the node is greater than the average energy of the network, then it has more chances to become a CH. Thus, energy is well distributed in the net work as it evolves.

\begin{figure*}[t]
\centering
\includegraphics [height=10cm,width=10cm]{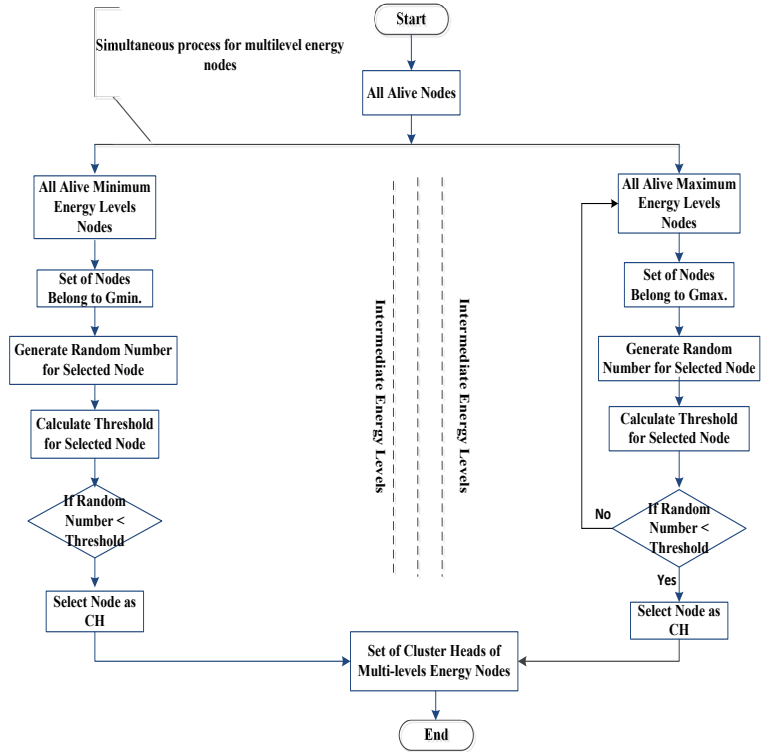}
\vspace{-0.4cm}
\caption{Flow chart of CH Selection in DEEC protocol.}
\end{figure*}

CH selection process of DEEC is give in a flow chart; Fig. 3. Table. I, briefly describes LEACH, TEEN, SEP, and, DEEC, protocols with respect to their mode of functioning and energy heterogeneity.

\begin{table*}[t]
%\caption {WSN Protocols and their features}
\begin {center}
\begin{tabular}{|l|l|p{4cm}|l|l|p{10cm}|} 
\multicolumn{5}{c}{Table.III. WSNs' Protocols and their features} \\
\hline
\textbf{Protocol}&\textbf{Routing Type}&\textbf{Initial Energy of Nodes} &\textbf{Hierarchal Level}&\textbf{CH Selection}\\ \hline
LEACH&Proactive&Homogenous&Single&Threshold-based Probability\\ \hline
TEEN&Reactive&Homogenous&Single&Threshold-based Probability\\ \hline
SEP&Proactive&Heterogenous&Bi-Level&Threshold-based on Weighted Probability\\ \hline
DEEC&Proactive&Heterogenous&Single&Residual Energy + Average energy of the Network\\ \hline
\end{tabular}
\end{center}
\end{table*}

\subsection{Cluster Formation Process in Hierarchical Clustering Protocols}
According to the hierarchical levels, The cluster formation process of LEACH, SEP and DEEC protocols is different from the TEEN. That is why, we characterize LEACH, SEP and DEEC as mono-level hierarchal protocols, whereas, TEEN is considered as mono, bi or multi-level hierarchical protocol i.e., hierarchy level depends upon network size.

\begin{figure}[h]
\centering
\includegraphics [height=9cm,width=9cm]{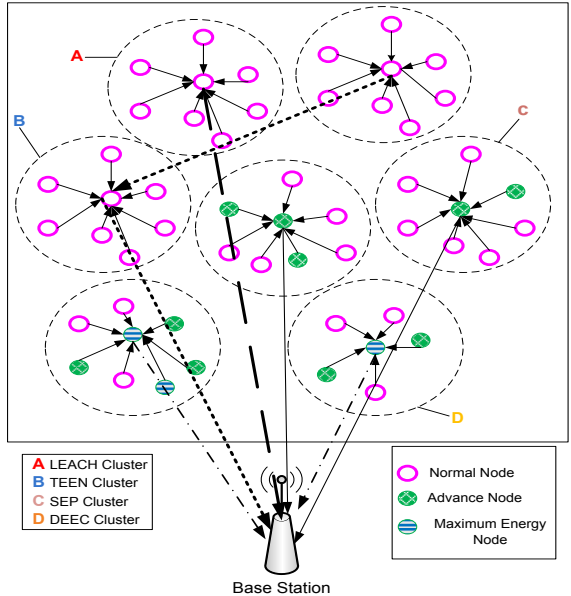}
\caption{Flow chart of basic mechanism in LEACH, TEEN, SEP and DEEC}
\end{figure}

\subsubsection{Cluster Formation Process in Mono-level Hierarchical Routing Protocols}
In LEACH, SEP and DEEC protocols, the cluster members nodes sense the required information from the environment in which they are deployed and then transfer their sensed information to the CH in the allocated time slots. The CH accumulates the cluster members sensed information and after evaluating compression on the sensed data, CH further transmits aggregated data to the BS. In the same manner, other clusters in these protocols, transmits their information to the BS, and thus, they make mono-level hierarchy in WSNs. The mono-level clustering hierarchy in LEACH, SEP and DEEC Protocols is also shown in Fig. 4.

\subsubsection{Cluster Formation Process in Multi-levels Hierarchical Routing Protocol}
TEEN further enhances LEACH by forming hierarchical levels of CHs. BS is at the top level in this hierarchy.  In this scheme CHs do not send their data directly to a BS, instead lowest level CH send its data to CH one level above in hierarchy and so on. This is the way, a TEEN protocol makes hierarchy in the network. In this way farther CHs save their energy by sending data to nearest CH. Although this scheme is beneficial for farther CHs, but not suitable for CHs that are near to BS. The multi-levels clustering hierarchy in TEEN protocol is also shown in the Fig. 4.

%\begin{figure}[h]
%\centering
%\includegraphics [height=7cm,width=8cm]{TEENhierarchical.eps}
%\caption{Multi-levels Clustering Hierarchy in TEEN Protocol.}
%\end{figure}

\section{Simulation Results}
In order to evaluate the performance of the selected protocols against maximizing objective function $K$, we perform analytical simulations in MATLAB. Simulation parameters are given in table II. For $N$, $100$ nodes are randomly scattered in network field of $100m\time 100m$ area. BS is placed at the center of the network field. In order to obtain more realistic results, we adjust the heterogeneity level for different routing protocols according to their proposed model. For energy dissipation characteristics, we adopted first order Radio Model, which is given in [1]. Before discussion of simulation results, it is necessary to define performance metrics. We will use following performance metrics in our results discussion.

\begin{table}[h]
%\caption {Simulation parameters}
\begin {center}
\begin{tabular}{|p{3.25cm}|p{3.25cm}|}
\multicolumn{2}{c}{Table.II. Simulation Parameters} \\
\hline
\textbf{Parameter}&\textbf{Value} \\ \hline
 Network size&$100\times 100$ meters\\ \hline
Minimum initial energy&$E=0.5 Joule$ \\ \hline
 $P_{opt}$&0.1 \\ \hline
Packet size&4000 bits\\ \hline
Transmit/ Receive Electronics&$E_{elc}=50 nJ/bit$ \\ \hline
Data Accumulation&$E_{DA}=5 nJ/bit/report$  \\ \hline
Transmitter Amplification $\left ( d\leq d_{0} \right )$&$E_{fs} =10 pJ/bit/m^{2}$ \\ \hline
Transmitter Amplification $\left ( d\geq d_{0} \right )$&$E_{mp} =0.0013 pJ/bit/m^{4}$ \\ \hline
\end{tabular}
\end{center}
\end{table}

\textbf{Stability Period:} Time duration between the starting of network process and expiry of very first node in the network.

\textbf{Instability Period:} Time duration between the expiry of very first sensor node and very last sensor node of the network.

\textbf{Network lifetime, $K$,:} Time duration between the network process initialization and the expiry of the very last alive sensor node in network.

\textbf{Cluster heads per round:} These are the some percentage of the nodes, that collect the sensed information of their associated cluster members and directly send to BS.

\textbf{Alive nodes per round:} These are total number of nodes that have not till yet expended all of their energy.

\textbf{Packets to BS:} These are total data packets that are successfully sent from their CHs to the BS.

Figure 5 shows stability period and network life time for of the network for all routing protocols with respect to alive nodes in $r_k$ number of rounds. We can observe that stable period of LEACH is very short. Stability period of LEACH is almost $23\%$, $55\%$, $50\%$ less than SEP, DEEC and TEEN, respectively. Because, LEACH treats all nodes without energy discrimination therefore it looses full advantage of nodes that have more energy. While SEP treats all the nodes with initial energy discrimination, therefore, the stability period of SEP is more than LEACH. DEEC has almost $25\%$, $10\%$ longer stable period than SEP and TEEN, as depicted in Fig. 6. This is because of heterogeneity-awareness of DEEC, which provides feasible solution against eq. 1(a-d). TEEN has better stability, as compare to LEACH and SEP. Reason behind this performance of TEEN is less number of transmissions, done by TEEN. Network life time of TEEN is almost $48\%$, $85\%$, $98\%$ greater than DEEC, SEP and LEACH, respectively.

\begin{figure}[h]
\centering
\includegraphics [height=5cm,width=8cm]{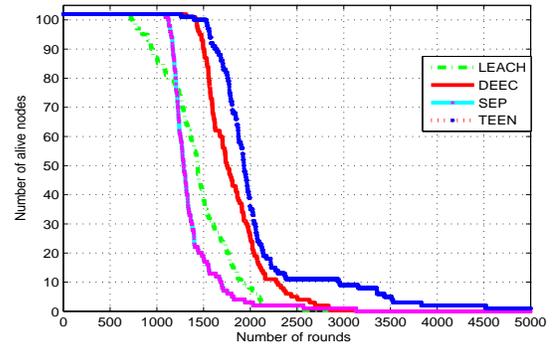}
\caption{Network Life Time of four Protocols}
\end{figure}

Figure. 6 shows number of dead nodes as network operation proceeds. Results shows the instability and value of $K$ of the network. We can see that network lifetime results are identical, as shown in previous Fig. 5. An important information that we can derive from this figure is instability faced by routing protocols that SEP has minimum and TEEN has maximum unstable region.

\begin{figure}[h]
\centering
\includegraphics [height=5cm,width=8cm]{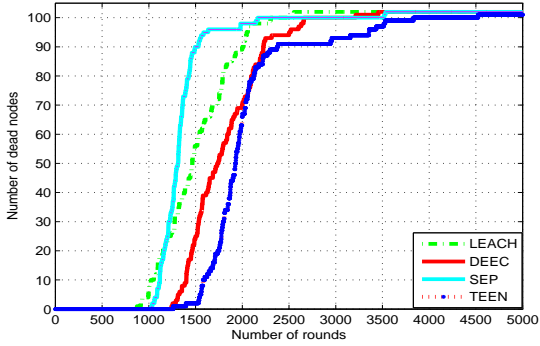}
\caption{Dead Nodes Versus Time.}
\end{figure}

Successful data delivery at BS is an important factor to analyze quality of routing protocol and it depends upon value of $E$, $K$ and resultant values of $x_{izp}$ and $\alpha_{ikz}$ during network life time, as mentioned in eq. 1. If BS is receiving high data it means routing protocol is working properly. Fig. 7 shows the comparison of every protocol for number of packets that are sent to BS. Result shows that DEEC has highest successful data rate, as compare to other routing protocols. It is because of shorter value of $K$ (In fig. 5) in LEACH and SEP, as compare to DEEC. However, TEEN has better network life time, as compare to DEEC, however its execution provide low data delivery, as compare to DEEC. Reason behind this unusual result is limited transmissions of TEEN. DEEC, is time-based routing protocols and it has to transmit data continuously. While TEEN is threshold based and have limited information to share with BS.

\begin{figure}[h]
\centering
\includegraphics [height=5cm,width=8cm]{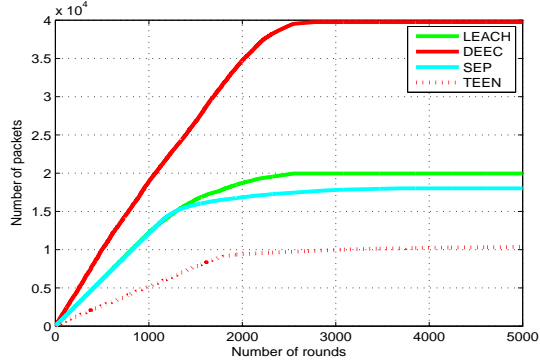}
\caption{Packets sent to the BS Versus Time (Rounds).}
\end{figure}
Figure. 8 depicts the number of CH which are selected in each round. These all routing protocols are utilizing distributed algorithm for selection of CHs. A main challenge faced by clustering routing protocols is their un-reliable distributed algorithm of selection of CHs as we formulate in eq. 1(b)(c). It is observed in Fig 8 that guaranteed number of CHs are not selected continuously. DEEC and TEEN mostly generate CHs above required average of CHs. Distributed algorithm generate un-even number of CHs for every round that can disturb performance of network where optimal number of CHs are necessary to enhance network's life.

\begin{figure}[h]
\centering
\includegraphics [height=5cm,width=8cm]{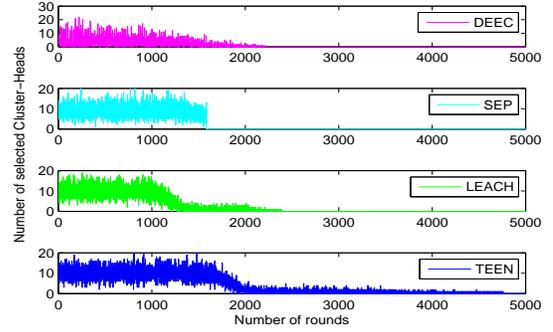}
\caption{Count of Cluster Head}
\end{figure}

The brief comparison of LEACH, TEEN, SEP and DEEC clustering protocols is given in table III. This analytical comparison clearly elaborates that LEACH and TEEN protocols support homogeneous environment and both have very limited scalability. SEP and DEEC are designed for heterogeneous network. These protocols are scalable than LEACH and TEEN protocols. TEEN protocol is highly energy efficient due to its event driven operation, therefore, it has highest value of $K$ as compared with other three protocols. SEP protocol is low energy efficient than the TEEN and DEEC protocols, however, it also has good life time than the LEACH protocol. DEEC protocol consumes less energy as compared to SEP and LEACH protocols, therefore, also has high value of $K$ than SEP and LEACH.

\section{Conclusion}
Energy optimization and efficient route discovery are challenging issues in WSNs. Different techniques have been proposed up till now to address these issues. Clustering technique is one of them, and this work is devoted to evaluate the efficiency of different clustering schemes. For this purpose, we first address energy limitation constraint with respect to maximizing network using linear programming formulation technique. To check the feasibility of different clustering techniques against modeled framework, we select LEACH, TEEN, SEP and DEEC. It is concluded from our analytical simulation results that DEEC is the most energy efficient protocol for heterogeneous node energy network.  However, TEEN is more energy efficient and attain highest value of $K$ due to its hard and soft threshold based communication. The energy consumption of TEEN is better than others due to its less data transmission to BS. Whereas, DEEC is efficient in sending maximum information to BS, while TEEN lacks due to its restriction on communication. SEP is good in selection of optimum number of CHs, and therefore produces small variations in CH selection. Thus overall DEEC outperforms among selected protocols by providing feasible optimum solutions against constraints of modeled frame work.

\end{document}